\documentclass{aastex}          
\usepackage{spr-astr-addons}    
\usepackage{cases}

\newcommand{\etal}{\it et al.}
\begin{document}
%

\title{North-south asymmetry in 
   small and large  sunspot group activity
and violation of  even-odd solar cycle rule}


\shorttitle{N-S asymmetry in small and large sunspot groups}
\shortauthors{J. Javaraiah}

\author{J. Javaraiah\altaffilmark{1}}

\email{jajj55@yahoo.co.in} 

\altaffiltext{1}{
\#58, BSK 5th Stage, Bikasipura (BDA Layout), Bengaluru-560 111, India \\
{Formerly  with Indian Institute of Astrophysics, Bengaluru-560 034, 
India}}


\begin{abstract}
According to Gnevyshev-Ohl (G-O) rule an odd-numbered 
cycle is stronger than its preceding even-numbered cycle.
In the modern time  the cycle pair~(22, 23) violated this rule.
By using the combined Greenwich Photoheliographic Results (GPR) 
and Solar Optical Observing Network (SOON)
 sunspot group data during the period 1874\,--\,2015, and Debrecen 
Photoheliographic Data (DPD) of sunspot groups during the period 1974\,--\,2015, 
 here we have found that the solar cycle pair (22, 23) 
violated the G-O rule because, 
 besides during cycle~23 a large deficiency of small sunspot  groups  in both 
the northern and the southern hemispheres,   during cycle~22 a large 
abundance of small sunspot groups in the southern hemisphere.
 In the case of large and small  sunspot groups the cycle pair (22, 23) 
 violated the  G-O rule in the  northern and southern hemispheres,
 respectively, suggesting the  north-south asymmetry in solar activity
  has a  significant contribution in the violation of G-O rule.  
The amplitude of solar cycle~24 is smaller than that of solar cycle~23.
 However, Coronal Mass Ejections (CMEs) rate in the rising phases of
 the cycles~23 and 24 are almost same (even slightly large in cycle~24).
From both the SOON and the DPD sunspot group data here we have also found
 that on the  average the ratio of the number (counts) of large
sunspot groups to the number of small sunspot groups is  larger in the
rising phase of cycle~24 than that in the corresponding phase of cycle~23.
We suggest this could be a potential reason for the aforesaid discrepancy
in the  CME rates during the rising phases of cycles 23 and 24.
These results have significant implication on solar cycle mechanism.
\end{abstract}


\section{Introduction}
A number of authors have shown the existence 
of a difference in the  number as well as the  
dynamic behaviors
of sunspots  in the northern and the southern
 hemispheres. 
 In fact, the existence of north-south asymmetry in solar 
activity is well established~\citep[][and references therein]{roy77,hath15}
 and the existence of several short- and long-term 
periodicities in the north-south asymmetry  is also 
known~\citep[][]{swin86,cob93,verma93,dd96,jg97a,li02,knaack04,chang09,chowd13,rj15}. 
 Phase relationship   
of activity in northern and southern
 hemispheres and its implication on Gnevyshev gaps, etc. were also 
investigated~\citep{tem06,zp06,dth07,ng10}. North-south asymmetry 
in the solar activity during  a solar cycle can be used to predict  
 amplitude of next cycle~\citep{jj07,jj08,jj15}. 
 Solar activity influences on the space weather and also may have an
 influence on the terrestrial 
climate~\citep{shap11,clette14,hath15,gopal15a}. 
North-south asymmetry of solar activity seems to have a role on 
  the atmospheric circulations~\citep{georg07}. 

One of the well known properties of solar cycles is the existence of 
  Gnevyshev-Ohl rule or G-O rule~\citep{go48}.
According to this rule 
an odd numbered cycle is stronger than the preceding even numbered cycle.
 However, some pairs of  even and odd numbered solar cycles violated this
  even-odd cycle rule. It seems the violation of  this rule is followed 
by a few weak cycles.  The cycle pair (22, 23) violated the even-odd 
cycle rule. The current  activity level is much lower in the last
 100 years also.   The activity trend over last 20 years  may  
 resemble the  Dalton minimum~\citep{zp14,jj15}.
  \cite{tltv15}  found that the secular minima 
of the solar activity occurs in the vicinity of the extreme points of the 
200-year cycle of inversion of G-O rule.

The dynamic behaviors 
of the  magnetic structures 
of large and small sunspot groups are different \citep{war65,war66},
 may be due to differences
 in dynamics of solar plasma    at  different subsurface layers of the
 Sun~\citep[$e.g.$][]{how96,jg97b,hi02,siva03,jj13}.
Recently it has been shown that  different classes of sunspot groups  behave
 differently over a cycle~\citep{klck11,lc11,clette12,jj12a,ob14}.
 In the earlier analysis~\citep{jj12a}
 it was found that the  cycle pair (22, 23)  violated  G-O rule in $R_{\rm Z}$
 due to a large deficiency of the small sunspot groups in cycle~23. 
In this paper we have investigated the
 north-south differences  in  the numbers of  large and small
 sunspot groups and their implication on the G-O rule. 
Such a study   is  important 
for understanding the  north-south asymmetry of solar activity 
relationship with the Sun's subsurface dynamics and solar cycle. 

The  solar flares and CMEs originate
from  solar active regions. Thus,  the frequencies of occurrence of CMEs
well correlated to the international sunspot number ($R_{\rm Z}$).
 The X-class flares occur in any phase of solar cycles~\citep{hath15}.
It is known that the CME productivity increases with active
region size~\citep{can99,ramesh10}. Recently, \cite{gopal15a} and
 \cite{gopal15b} have  found that though the peek of the
 current sunspot cycle~24 is smaller than that of sunspot cycle~23, in the
 rising phase of cycle~24  halo CMEs abundance  is relatively large. 
This discrepancy in the behavior of CME rates and sunspot activity in the
rising phases of cycles~23 and 24 is not yet understood~\citep{gopal15b}.
 In view of
the aforesaid discrepancy in CME rates during the rising phases of
the  cycles 23 and 24,
 here we have also investigated on the ratio of the number of
large sunspot groups  to the number of small sunspot groups  during the
rising phases of cycles 23 and 24, because it may be providing an
important clue for understanding
 the aforesaid  discrepancy in the CME rates.

In the next section we  described the data analysis, in Section~3 we  
 described the results, and in Section~4 we  summarized the results  
 and  their implications.

\section{Data analysis}
Here we have used the combined Greenwich and SOON sunspot group 
data during the period 1874\,--\,2015 (available at 
{\tt http://solarcience.msfc.nasa.gov\break /greenwich.shtml}).
David Hathaway  scrutinized
the GPR and SOON sunspot group  data and produced a reliable
continuous data series
from 1874 up to date~\citep{hath03,hath08,hath15}.
The Royal Greenwich Observatory terminated the
 publication of GPR
at the end of 1976. Since 1977 Debrecen Heliophysical Observatory
took over this task~\citep[for detail see][]{gyr10}. The DPD sunspot
 group data during 1974\,--\,2015  are available at 
{\tt http://fenyi.solarobs.unideb.hu/pub/DPD/}. We also analyzed this data 
and compared the results found from this and SOON data sets. 
The data reduction and analysis are same as in \cite{jj12a}. 
The data consist the values of the 
date and the time of observation, 
heliographic latitude and longitude, corrected whole spot area ($A$), 
etc., of each of the sunspot groups observed in each day during their
 respective life times of the sunspot groups.  
In case of SOON data, we increased area by a factor of 1.4. This 
is necessary to have a uniform combined GPR and SOON data~\citep{hath08,hath15}. 
The maximum area of a sunspot group having life time $n$ days 
  is defined as $A_{\rm M} = {\rm max} (A_1, A_2,...,A_n$),
 where $A_1$, $A_2$,...,$A_n$ are area measured in $
1^{st}$, $2^{nd}$, $\dots$, $n^{th}$ days.
We have used here only the sunspot groups having life time 
at least $n = 2$ days.
 In the aforementioned 
paper  relatively long-term variations in the yearly numbers
 (counts) of the small
 (maximum area $A_{\rm M} < 100$ millionth of solar hemisphere, msh),
 large ($100 \le A_{\rm M} < 300$ msh),
and very-large ($A_{\rm M} \ge 300$ msh) sunspot groups are studied by
 analyzing the combined data of all the sunspot groups occurred in
 the Sun's whole sphere (Note: here we have replaced the word `big', which was 
used in our earlier papers, with  words `very-large'. This is because we
 find that
the  words `large' and `big' give same/equivalent meaning. Hence, it causes
 some confusion 
to the readers on the classifications of the sunspot groups).
 The   north-south difference/asymmetry in the solar cycle variations 
 in the yearly  numbers of 
 the small, the large, and the very-large sunspot groups 
are studied. The ratio of the number of large to the number of small sunspot
groups during the ascending phases of solar cycles~23 and 24 are also determined
(note: for the sake of better statistics, in this case the large and the
 very large sunspot groups have been combined.)

\section{Results}

\begin{figure}
\centering
\includegraphics[width=8.0cm]{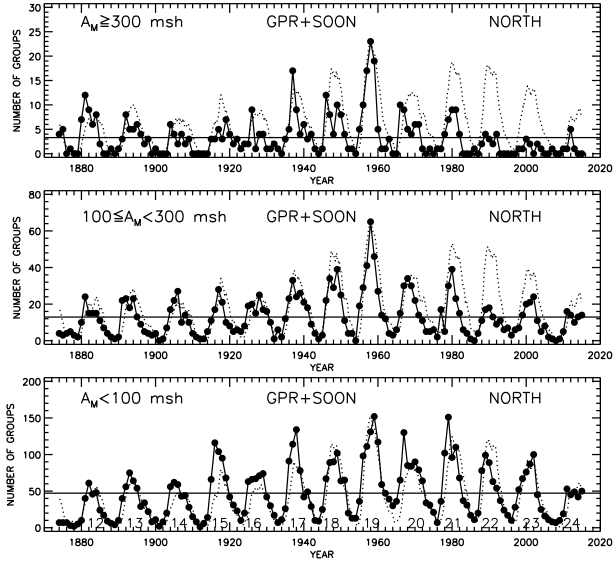}
\caption{Plots of the values of the numbers of  small (lower panel), 
large (middle panel), and very-large (upper panel) sunspot groups
  in the northern hemisphere versus time (year) during the period
 1874\,--\,2015. The horizontal lines represent the corresponding mean values.
The dotted curve represents variations in the  normalized 
13-month smoothed $R_{\rm Z}$. 
In the lower panel near maximum epoch of each solar cycle the corresponding
Waldmeier cycle Number is given.}
\label{f1}
\end{figure}

\begin{figure}
\centering
\includegraphics[width=8.0cm]{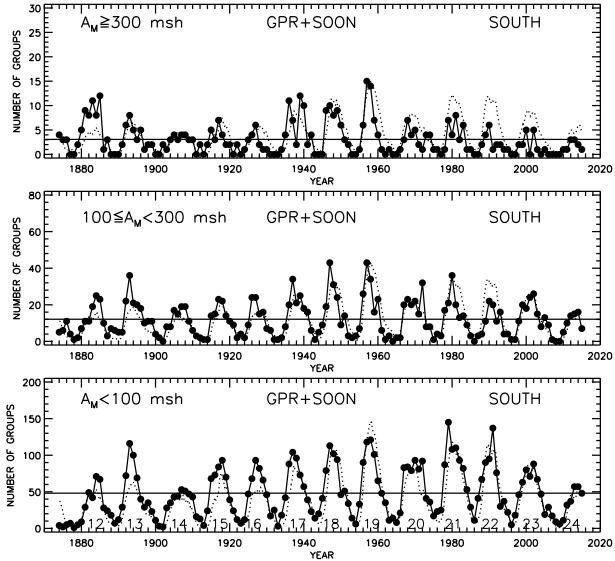}
\caption{The same as Fig.~\ref{f1} but determined from the data of  the sunspot
 groups in the southern hemisphere.}
\label{f2}
\end{figure}

\begin{figure}
\centering
\includegraphics[width=8.0cm]{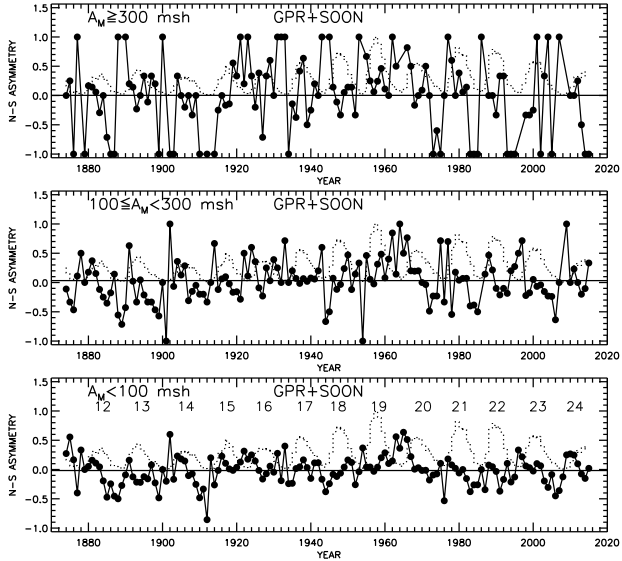}
\caption{\bf Plots of the values of the north-south asymmetries in the
 numbers of  small (lower panel), large (middle panel), and
 very-large (upper panel) sunspot groups  versus time (year) during the period
 1874\,--\,2014. The horizontal lines represent the corresponding mean values.
The dotted curve represents variations in the normalized 13-month smoothed 
$R_{\rm Z}$. In the lower panel near maximum epoch of each solar cycle 
the corresponding Waldmeier cycle Number is given.}
\label{f3}
\end{figure}

Figs.~\ref{f1} and \ref{f2} show variations in the yearly numbers of small,
 large,
 and very-large sunspot groups in the northern and the southern hemispheres, 
 respectively, during the period  1874\,--\,2015. 
 For the sake of checking the solar cycle 
properties in these variations,  in Figs.~\ref{f1} and \ref{f2} we have also
 showed  
the  variation in the 13-month smoothed  international sunspot number
 $R_{\rm Z}$, 
taken from {\tt http://www.ngdc.noaa.gov/} (Note: for the sake of scaling 
$R_{\rm Z}$ is normalized).  
 As can be seen in these figures  
 there exist 11-year solar cycle patterns in the variations
 of all the three classes of sunspot groups in  each hemisphere.  
Overall the variations  
 in the numbers of small, large, and very-large sunspot groups in  each
 hemisphere 
are largely same as the corresponding variations determined by  
  \cite{jj12a} from the whole sphere data.  
 However, the existence of north-south  
  differences in these  variations are noticeable  
during   some  cycles.

\begin{figure}
\centering
\includegraphics[width=8.0cm]{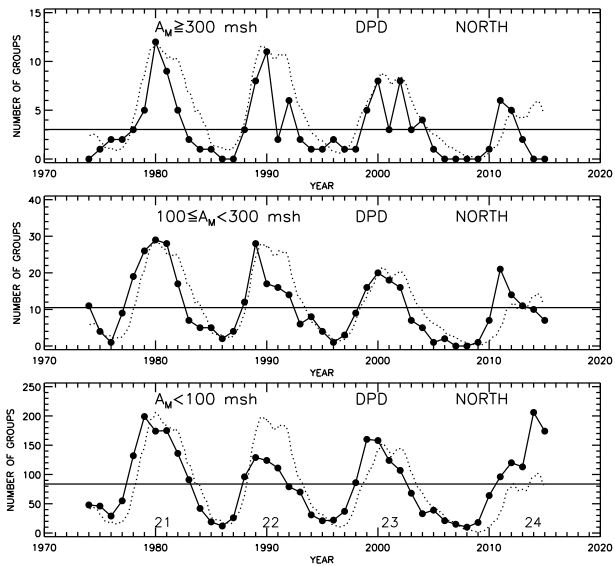}
\caption{The same as Fig.~\ref{f1} but determined from the DPD 
 data of  the sunspot groups  during 1974\,--\,1915.} 
\label{f4}
\end{figure}

\begin{figure}
\centering
\includegraphics[width=8.0cm]{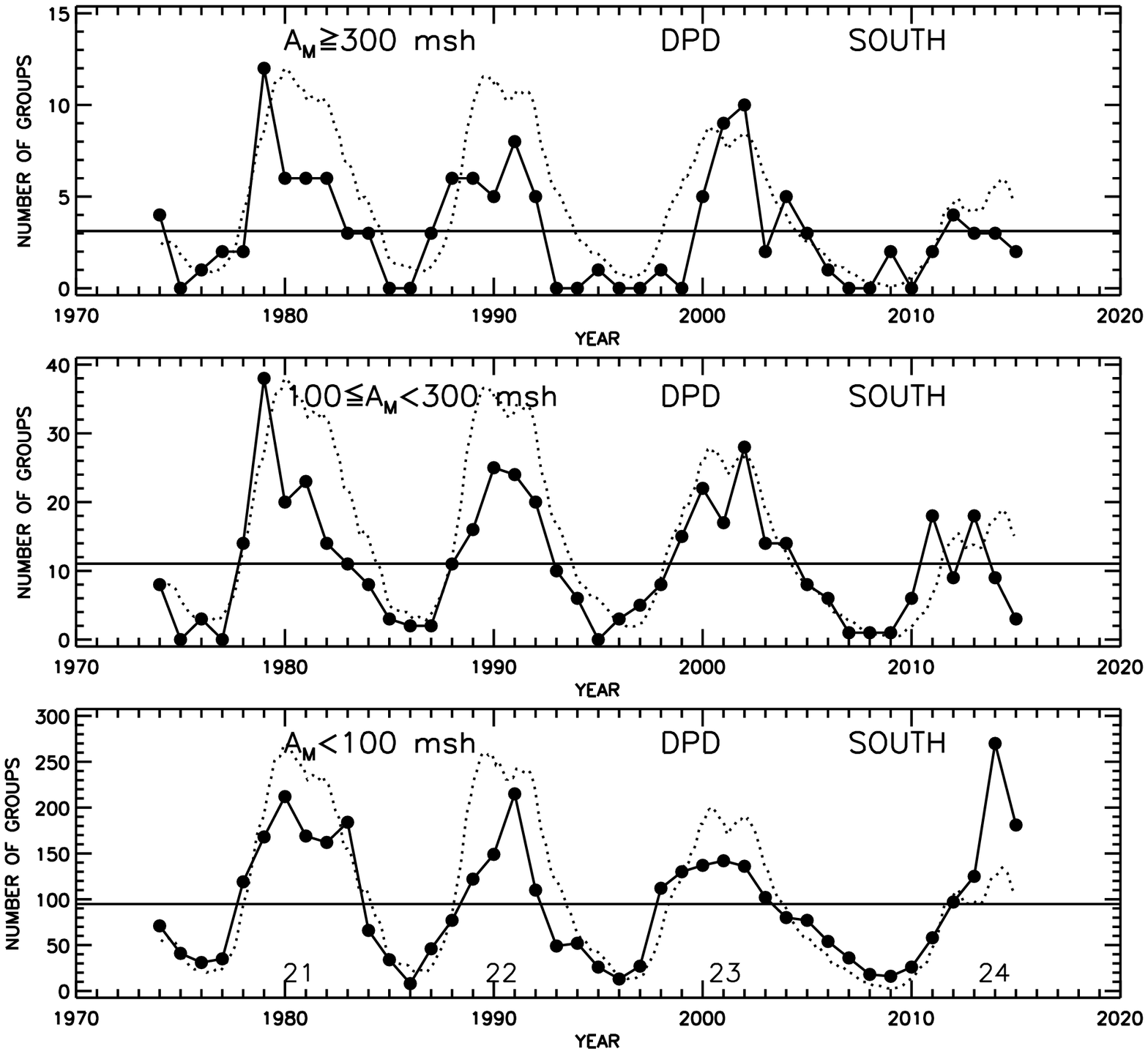}
\caption{The same as Fig.~\ref{f2} but determined from the DPD 
 data of  the sunspot groups  during 1974\,--\,1915.} 
\label{f5}
\end{figure}

\begin{figure}
\centering
\includegraphics[width=8.0cm]{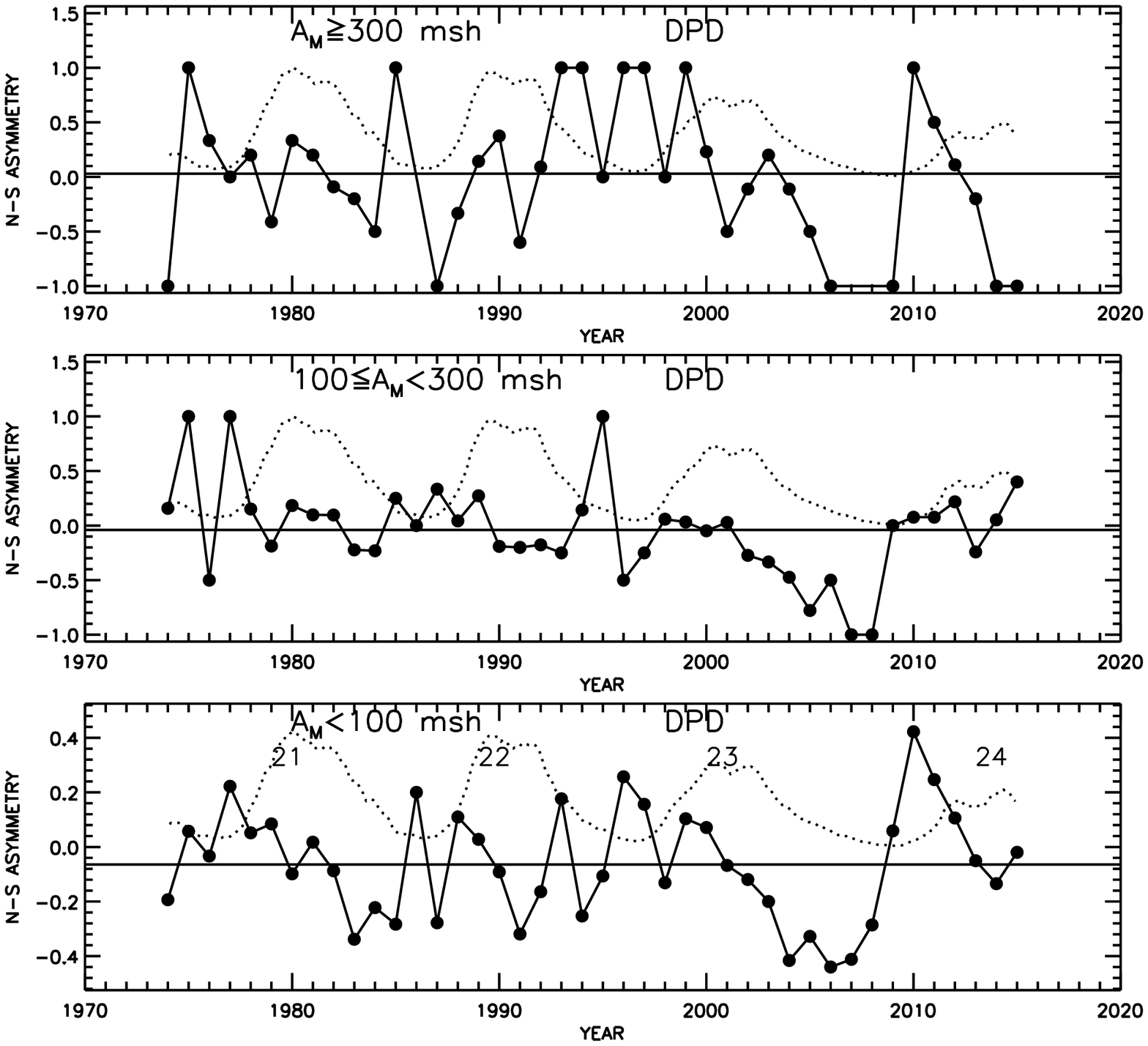}
\caption{The same as Fig.~\ref{f3} but determined from the DPD 
 data of  the sunspot groups  during 1974\,--\,1915.} 
\label{f6}
\end{figure}

In many studies  the north-south asymmetry in solar activity is determined 
  as $(N-S)/(N+S)$, where $N$ and $S$ are the amounts of  
 activity  in the Sun's northern and southern hemispheres, 
respectively~\citep{nm55,swin86,cob93,verma93,dd96,jg97a,knaack04}. 
Because of  relatively less error 
 in this ratio~\citep{jg97a}, it may be worth to check the variation 
in this ratio rather than in the absolute north-south difference. 
  Fig.~\ref{f3} shows
the variations in the corresponding north-south asymmetry  of the yearly
 counts of small, 
large, and very-large sunspot groups. In this figure we have also showed the 
 variation in the 13-month smoothed  $R_{\rm z}$. In this figure the extreme 
values $-1$ and $+1$ of the 
north-south asymmetry in the number of a  class of sunspot groups
imply the absence of the corresponding class of sunspot groups in
the northern and the southern 
 hemispheres, respectively. 
 Such epochs are more in the case of 
the number of very-large sunspot groups, causing  large inconsistency
 in the corresponding
 time series. In some years the values are equal to  zero because  
  the  numbers of corresponding sunspot groups are equal  in  northern 
and southern 
hemispheres. There are some  gaps in the time series (see closed circle curve)
due to absence of the data in both the hemispheres. Such gaps  
 are also more in the case of the number of 
very-large sunspot groups. 
As can be seen in this figure,  the   north-south 
asymmetry patterns in   the numbers of all the three classes of
 sunspot groups are 
 similar. There is also  an indication that the north-south asymmetry is 
 multiperiodic in nature. 
 In the declining phases closer to the minima of a large 
number of solar cycles the values of the north-south asymmetry are  
relatively  high in each class of sunspot groups.  
There are some   trends in the north-south asymmetry  of 
the number of large sunspot groups
(and even in the asymmetry of the number of  very-large sunspot groups) which 
suggest the following:  before  cycle~16 
 during the minima of  a large number of cycles 
large sunspot groups were more    
 in the  southern hemisphere,  
  during a large number of cycles between cycles~16 and 20 large 
sunspot groups were more in   
 the  northern hemisphere,   and after
cycle~20 during minima of the remaining   cycles there seems to be the 
 large sunspot groups were somewhat  more in the southern hemispheres.
 At maximum epoch of cycle~23   the north-south
 asymmetry is very small in the  number of 
small sunspot groups, whereas the asymmetry in the number of small 
sunspot groups  is reasonably 
large at the  maximum epoch of cycle~22  
 due  to a large number of small sunspot groups 
 in the southern hemisphere.
In the declining phase of the strongest cycle~19, the asymmetry in each of 
 the three classes of sunspot groups has positive values suggesting 
   sunspot activity is large in the 
northern hemisphere. The trends indicate the existence of 
  11\,--\,12 year periodicity in the north-south asymmetry in the   
numbers of both the large and the small sunspot groups.
 In fact, the existence 
of this periodicity in the north-south asymmetry of solar activity is 
 known~\citep{cob93,jg97a}.  
The patterns of the north-south asymmetry in the number of small 
sunspot groups (and some extent the pattern  in the number of large 
sunspot groups)  during cycles 12\,---\,14 and
 cycles 21\,--\,23 are similar and
they are  differing with 
the pattern during cycles 16\,--\,20. That is, on the average 
 over cycles 12\,--\,14 and also over cycle  21\,--\,23 the asymmetry
 seems to be negative (southern hemisphere 
dominance) and a large contributions to this property    
have come from the declining phases of these cycles. This long-term 
pattern suggests the existence of 55\,--\,65 year periodicity in the   
north-south asymmetry of the number of small sunspot groups.

As can be seen in Fig.~\ref{f1} in the case of the number of small
 sunspot groups 
in the northern hemisphere cycle~18 is much 
weaker than cycles~17, and the cycles~22 and 23 are approximately equal in 
strength. As can be seen in Fig.~\ref{f2} in the 
southern hemisphere cycle~18 is slightly stronger than cycle~17, and in 
fact, the strengths of the cycles~18 and 19 are approximately  equal. 
 As can be seen in Figs.~\ref{f1} and \ref{f2} 
in the case of small sunspot groups   
in the northern hemisphere 
the even-odd cycle rule  is not well defined  during cycle~22 and cycle~23,
 whereas  in the southern  hemisphere  
it is very clear  cycle pair (22, 23) violated the even-odd cycle rule 
 (the difference between the corresponding  amplitude of cycles~22 
and 23 is significant on $95\%$ confidence level).  In the case of large 
sunspot groups throughout cycles 12\,--\,23 the validity of even-odd cycle
 rule  is unambiguous. That is, this rule seems to be valid even in the case
 of cycle pair (22, 23) in  both the northern and the southern hemispheres,
 in consistent with the similar result found  from the whole sphere
 data~\citep{jj12a}. 
 In the earlier analysis~\citep{jj12a}
 it was found that the  cycle pair (22, 23)  violated  G-O rule in $R_{\rm Z}$
 due to a large deficiency of the small sunspot groups in cycle~23. 
 Further 
here we find that the  violation is caused mainly due to,
 besides in cycle~23 a large deficiency of small sunspot  groups in both 
 the northern and southern hemispheres, a large  
 abundance of small sunspot groups  during cycle~22 in the southern hemisphere.
This is consistent with the pattern of north-south asymmetry  in the 
number of small sunspot groups of cycles 22 and 23 as found above. 

Figs.~\ref{f4} and \ref{f5} show the variations in the numbers of small, large, 
and very-large sunspot groups in the northern and southern hemispheres, 
respectively, determined from the DPD sunspot group data during the 
period 1974\,--\,2015. Fig.~\ref{f6} shows the variations in the corresponding 
north-south asymmetry in each of the three classes of sunspot groups. 
There is a large agreement between the variations in the number of 
each class of sunspot groups  shown in these figures  
 with  the corresponding  variations during cycles 
21\,--\,24   shown in Figs.~\ref{f1}, \ref{f2}, and \ref{f3}.
 Overall   the results 
found above  from the SOON sunspot group data during cycle 21\,--\,24 
are  consistent  with the  
variations shown in Figs.~\ref{f4}, \ref{f5}, and \ref{f6}.    
 As can be seen in 
Figs.~\ref{f4} and \ref{f5} in the case of small sunspot groups the
 violation of G-O rule
by cycle pair (22, 23) is only in the southern hemisphere, 
whereas in the 
case of large sunspot groups it was happened only in northern hemisphere. 
 Overall  it seems north-south asymmetry   has a  
significant contribution in the violation of G-O rule.  
During the Maunder minimum the
 activity (small sunspot groups in low latitudes) was present mainly
 in the southern hemisphere~\citep{snr94}.
As can be seen Fig.\ref{f6} during the  prolonged deep minimum 
between cycles 23 and 24 the north-south asymmetry in the 
number of small sunspot groups  also 
indicates the southern hemisphere dominate with more small sunspot groups.

\cite{klck11} have 
 found that in general large sunspot groups  
peaked about two years later than the small ones.
We found that in many cycles the positions of the peaks of 
the small, large, and very-large sunspot groups  are
different, and they also deviate considerably from the corresponding peak
positions of $R_{\rm Z}$~\citep{jj12a}. 
The current sunspot cycle~24 has double peaks or Gnevyshev 
peaks~\citep{gnev63,gnev67}. 
 \cite{ng10} found  
 that the Gnevyshev Gap ($viz.$, the gap between the Gnevyshev peaks)
 is a phenomena that occurs
in both hemispheres and is not due to the superposition of two hemispheres
 out of phase with each other.
\cite{kz14}  suggested  that one possible reason for a double-peaked
 maximum in a solar cycle is the different behavior of large and small sunspot 
groups, resulting from the existence of two different dynamo mechanisms. 
That is, the double-peaked maxima of solar cycles may be caused by
 a bi-dynamo mechanism~\citep{du15}. 
The   second peak of cycle~24 that took place at the year 2014 is stronger
 than the first peak that took place at year 2012. 
It can be seen in Figs.~\ref{f4} and \ref{f5} 
 the second peak is dominant due to it consists of a large number of
 small sunspot groups, both in the northern and southern hemispheres. 
The first peak contains 
relatively more number of large sunspot groups.

(Note: Cycle~24 is the smallest solar cycle since cycle~14~\citep{sva05,du11,jj15}. As can be seen in Figs.~\ref{f4} and \ref{f5}
 although the current sunspot cycle~24 is weaker than previous 
sunspot cycle~23, 
the number of small sunspot groups at maximum (second peak) of cycle~24 
is larger than that at maxima of cycles~21\,--\,23, in both the northern and
 southern hemispheres.  
Figs.~\ref{f1} and \ref{f2} show that in cycle~24
 the  peaks of the small sunspot groups  are small in both the 
northern and southern hemispheres 
 and their heights close to the 
heights of the corresponding peaks of the weak cycle~14. 
That is, at maximum epoch of cycle~24 there exists a considerable 
difference between  SOON and DPD data. We don't know 
the reason behind this difference. However,  
generally at any time  small magnetic regions  dominate the large ones, and
the dominant variations in $R_{\rm Z}$ mostly depict the dominant 
variations in the number of small active regions.  Since the cycle~24 
is much weaker than cycle~23,  one can expect at maximum of
 cycle~24 the small sunspot groups should not exceed the small sunspot 
groups at maximum  of cycle~23. Therefore, at maximum of cycle~24
 the aforesaid behavior of SOON data may be correct.
In DPD data NOAA sunspot group number is assigned if it exists and it has not 
been revised. If no NOAA number was assigned for the
 group, a NOAA number was given with an additional
 letter (e.g. "m", "n", etc.). 
From our way of classification of sunspot groups on the 
basis of their maximum areas, we found that in the years 2014 and 2015 
we have got the large number of small sunspot groups due to the presence of 
 a large number of the daily data records of these years  with a NOAA number
 having the additional letters.)

\begin{figure}
\centerline{\includegraphics[width=8.0cm]{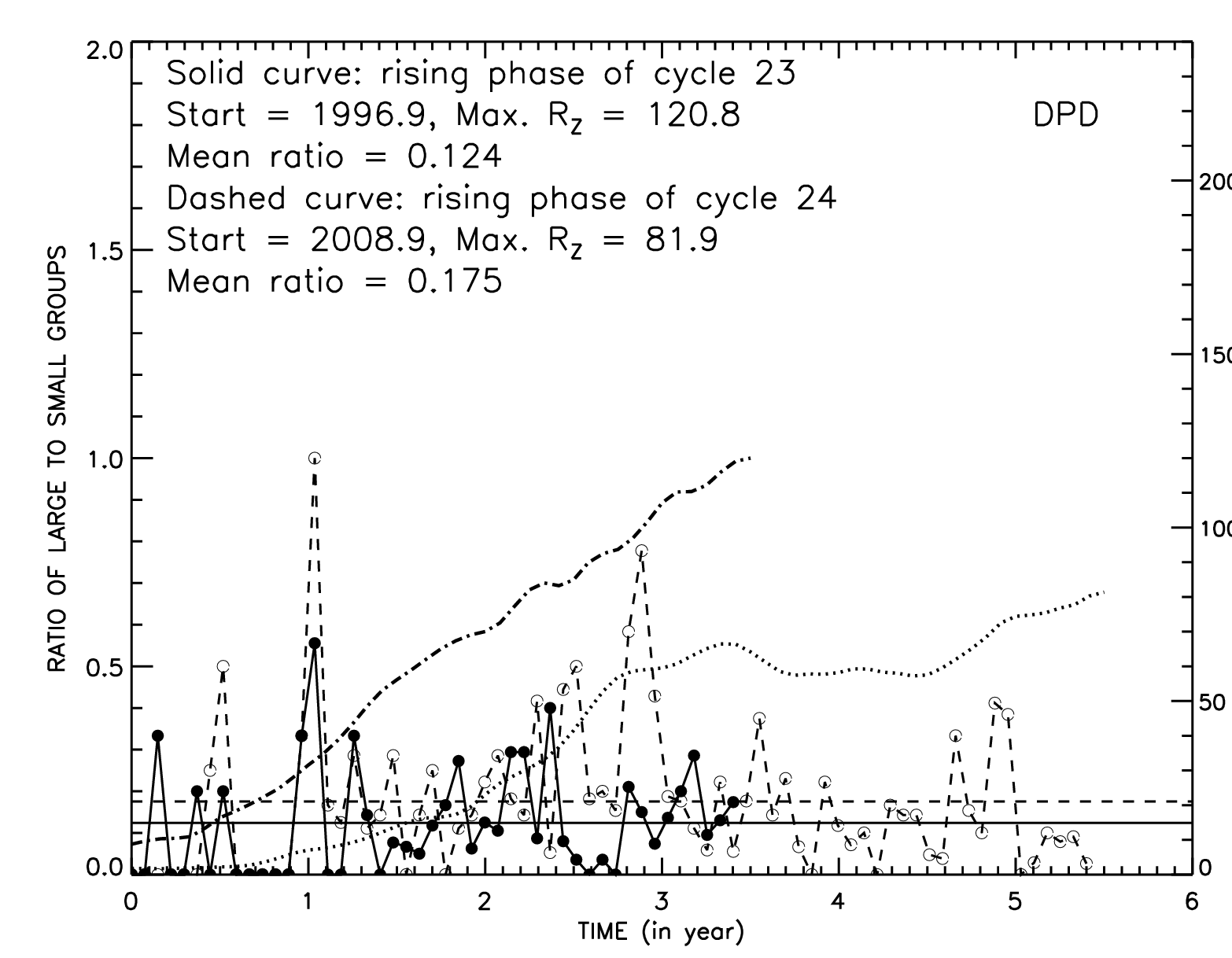}}
\centerline{\includegraphics[width=8.0cm]{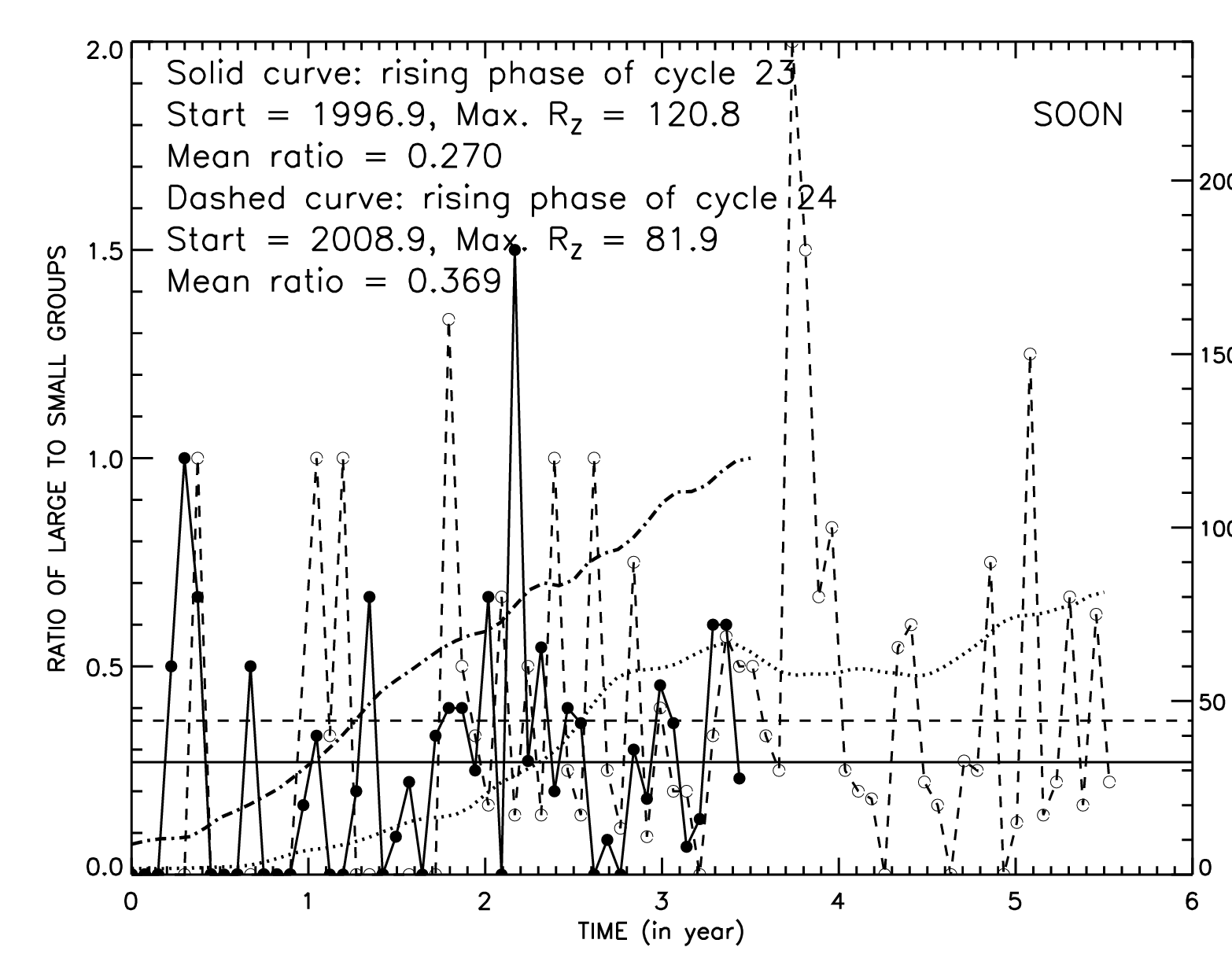}}
\caption{Plots of  the values of the ratios of the number of the  large
 sunspot groups to the number of the small sunspot groups
in the whole disk during n the rising
phases of solar cycles~23 and 24  versus time (in the intervals of 27-day), 
 determined from DPD (upper panel) and
SOON (lower panel) sunspot group data.  
(Note: the large and very-large sunspot groups have been combined.)
The horizontal lines represent the corresponding mean values.
The  dotted-dashed and  dotted
 curves represent the  variations in the 13-month smoothed monthly
 $R_{\rm Z}$ during the rising phases of cycles~23 and 24, respectively.} 
\label{f7}
\end{figure}

\begin{figure}
\centerline{\includegraphics[width=8.0cm]{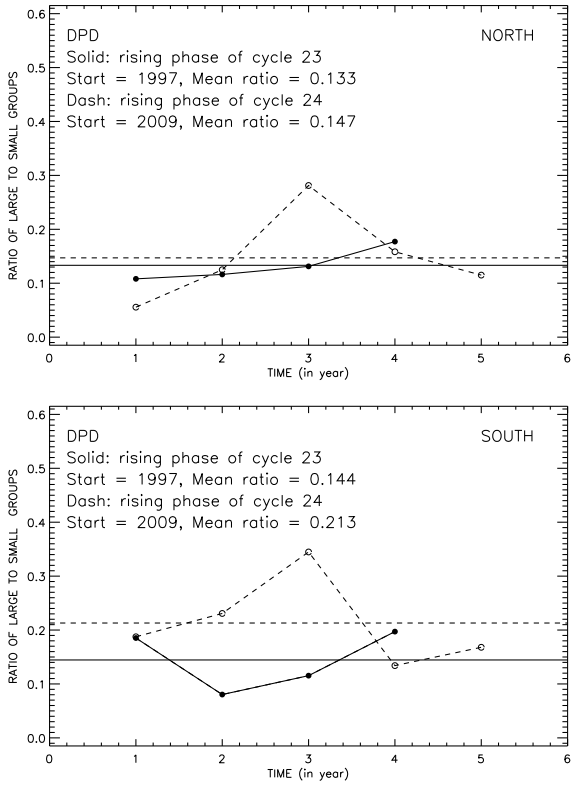}}
\caption{Plots of  the yearly values of the ratios of the numbers of the  large
 sunspot groups to the number of the small sunspot groups
 determined from the  northern hemisphere (upper panel) data 
 and the southern hemisphere (lower panel) data during
the rising phases of the cycles' 23 and 24  versus time (year).
(Note: the large and very-large sunspot groups have been combined.)
The horizontal lines represent the corresponding mean values.}
\label{f8}
\end{figure}

\begin{figure}
\centerline{\includegraphics[width=8.0cm]{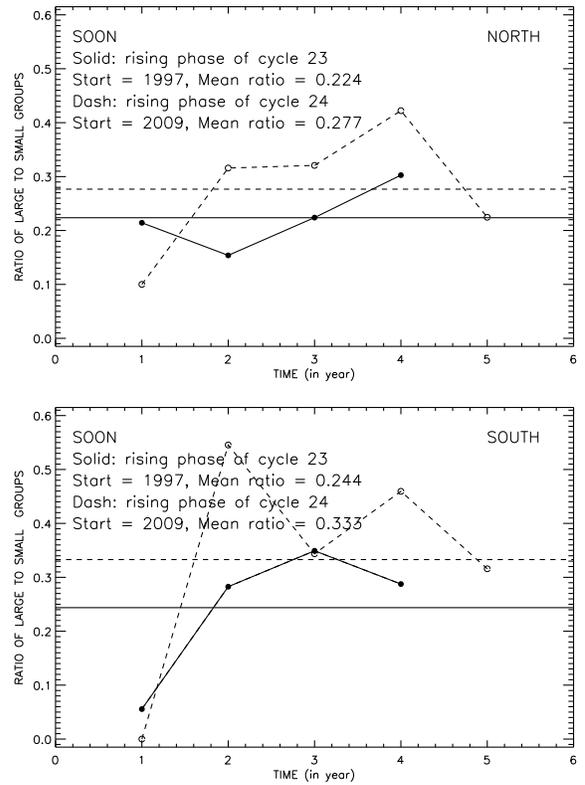}}
\caption{The same as Fig.~\ref{f8} but determined from SOON sunspot group data.}
\label{f9}
\end{figure}

Fig.~\ref{f7} shows variations in   the ratio of the number
of large sunspot groups to the number of small sunspot groups in 27-day 
consecutive intervals during   the  rising phases of solar cycles~23 and 24. 
\cite{gopal15a} and \cite{gopal15b} studied the CME  variations in 27-day
 intervals.
In order to check whether the large to 
small sunspot group ratio match with that of CME variation shown in Fig. 7 
of \cite{gopal15a}  here we have also used 27-day intervals.
 Here we have also combined the 
large and very-large sunspot groups in order to have a better statistics
(Note: the pattern of this combination is highly similar to that of the number 
of large sunspot groups.)  It is found that there 
are  some differences and some similarities within 
 the patterns of the CME and the ratio of the  sunspot groups. 
 Figs.~\ref{f8} and \ref{f9} show the variations in the ratio of the
 number of large to the number of small sunspot groups 
 in the northern and southern hemispheres. In this case  for the sake of
better statistics we have used yearly data.
 As can be seen in these  figures there are considerable differences in the
 corresponding variations determined from the SOON and DPD data sets.
(The sunspot data from  different observatories yield the results which
are generally differ with 5\%-10\% percent). 
 As can be seen in Fig~\ref{f7}, the results
determined from both the SOON and DPD data suggest that
 during the rising phase of cycle~24  in many places
 the values of the ratios of the number of large  to the number of
small sunspot
 groups are larger than the corresponding values during the rising
phase of cycle~23. Hence, the average values of
 the  ratios in the rising phase of cycle~24  is considerably
larger than the corresponding average values during  the rising phase of
 cycle~23. A similar property can also be seen in
 Figs.~\ref{f8} and \ref{f9}. The difference between the mean values of
the ratios of cycles~23 and 24  seem
 to be slightly larger in the southern hemisphere than  in the northern
 hemisphere.

\section{Conclusion and discussion}
From the analysis and the results above we can draw the following conclusions:
\begin{enumerate}
\item The solar cycle pair (22, 23) violated the 
  G-O rule of sunspot cycles  mainly due to,
 besides in cycle~23 a large deficiency of small sunspot  groups  in both 
the northern and southern hemispheres, during cycle~22 a large 
abundance of small sunspot groups in the southern hemisphere.
\item In the case of large and small  sunspot groups the cycle pair (22, 23)
 violated the  G-O rule in the  northern and the hemispheres,
 respectively, suggesting the north-south asymmetry in solar activity 
  has a  significant contribution in the violation of G-O rule.  
\item In both the northern and  southern hemispheres the average ratio of
 the number of large to the number of small sunspot groups  is
larger in the rising phase of cycle~24 than that in the same phase of cycle~23.
This could be a reason behind the CMEs (halo) are  more abundant
(in spite of the low sunspot activity) in the rising phase of cycle~24 than
 in the same  phase of cycle~23. 
\end{enumerate}

The  mechanisms  of the generations of the magnetic structures of 
the  large and the small sunspot groups 
may be associated with plasma dynamics  at  deeper and sallower 
 layers, respectively, of the Sun's convection 
zone~\citep[][and references therein]{jg97b,jj13}. That is,  
the magnetic structures
of large sunspot groups are deep rooted than the 
those of small sunspot groups.
This could be 
 responsible for  the  above said difference in the north-south
 asymmetry of the numbers of  large sunspot groups  with that of 
 small sunspot groups. The  conclusion 3 above is consistent with  
the  known result that the CME productivity increases with active
region size~\citep{can99,ramesh10}.  
That is, a large sunspot group
 could produce  relatively a large number of CMEs. 
Large sunspot groups also live long and their rate
 of evolution/decay also seem to be relatively large~\citep{jj11,jj12b}.
 Hence, during the evolution/decay of a large sunspot group  the release of
 relatively large amount of underneath stored
 thermal energy may be  responsible for more CMEs.

\acknowledgments
The author thanks the anonymous referee for useful comments and suggestions.

{}

\begin{thebibliography}{}
\bibitem[\protect\citeauthoryear{Canfield \etal}{1999}]{can99}
Canfield, R.C., Hudson, H.S.,  McKenzie, D.: \grl\ {\bf 26}, 627 (1999)
\bibitem[\protect\citeauthoryear{Carbonell \etal}{1993}]{cob93}
Carbonell, M., Oliver, R., Ballester, J.L.: \aap\ {\bf 274}, 497 (1993)
\bibitem[\protect\citeauthoryear{Chang}{2009}]{chang09}
Chang, H.-Y.: \na\ {\bf 14}, 133 (2009)  
\bibitem[\protect\citeauthoryear{Chowdhury \etal}{2013}]{chowd13}
Chowdhury, P., Choudhary, D. P., Gosain, S.: \apj\ {\bf 768}, 188 (2013)
\bibitem[\protect\citeauthoryear{Clette and Lef\`evre}{2012}]{clette12}
Clette, F., Lef\`evre, L.: J. Space Weather Space Clim. {\bf 2}, A06 (2012) 
\bibitem[\protect\citeauthoryear{Clette \etal}{2014}]{clette14}
Clette, F., Svalgaard, L., Vaquero, J.M., Claver, E.W.: \ssr\ {\bf 186}, 35 (2014)
\bibitem[\protect\citeauthoryear{Donner and Theil}{2007}]{dth07}
Donner, R., Thiel, M.: \aap\ {\bf 475}, L33 (2007)
\bibitem[\protect\citeauthoryear{Du}{2015}]{du15}
Du, Z.L.: \apj\ {\bf 804}, 3 (2015) 
\bibitem[\protect\citeauthoryear{Du and Wang}{2011}]{du11}
Du, Z.L., Wang, H.N.: Res. Astron. Astrophys. {\bf 11}, 1482 (2011) 
\bibitem[\protect\citeauthoryear{Duchlev and Dermendjiev}{1996}]{dd96}
Duchlev, P.I., Dermendjiev, V.N.: 1996: \solphys\ {\bf 168}, 205 (1996)
\bibitem[\protect\citeauthoryear{Georgieva \etal}{2007}]{georg07}
Georgieva, K., Kirov, B., Toney, P., Guineva, V., Atanosov, D.:  
  Adv. Space Res.  {\bf  40}, 1152 (2007) 
\bibitem[\protect\citeauthoryear{Gnevyshev}{1963}]{gnev63}
Gnevyshev, M.N.: 1963, \sovast\ {\bf 7}, 311 
\bibitem[\protect\citeauthoryear{Gnevyshev}{1967}]{gnev67}
Gnevyshev, M.N.: \solphys\ {\bf 1}, 107 (1967) 
\bibitem[\protect\citeauthoryear{Gnevyshev and Ohl}{1948}]{go48}
Gnevyshev, M.N.,  Ohl, A.I.: \azh\   {\bf 25}, 18 (1948)
\bibitem[\protect\citeauthoryear{Gopalsawmy \etal}{2015a}]{gopal15a}
Gopalswamy, N., Tsurautani, B.,  Yan, Y.: Prog. Ear. Plan. Sci. {\bf 2}, 13
 (2015a)
\bibitem[\protect\citeauthoryear{Gopalswamy \etal}{2015b}]{gopal15b}
Gopalswamy, N., Xie, H., Akiyama, S., M\"akela, P., Yashiro, S.,
Michalek, G.: \apjl\  {\bf 804}, L23 (2015b).
\bibitem[\protect\citeauthoryear{Gy\"ori \etal}{2010}]{gyr10}
Gy\"ori, L., Baranyi, T.,  Ludm\'any, A.:  Proc. Intern. Astron.
Union 6, Sympo. S273,  {\bf 2011}, 403 (2010) DOI: 10.1017/s174392131101564X
\bibitem[\protect\citeauthoryear{{Hathaway}}{2015}]{hath15}
Hathaway, D.H.:  Living Rev. Sol. Phys. {\bf 12}, No.4 (2015)
 (arXivr,150207020v1)
\bibitem[\protect\citeauthoryear{Hathaway and Choudhary}{2008}]{hath08} 
Hathaway, D.H., Choudhary, D.P.: \solphys\ {\bf 250}, 269 (2008)
\bibitem[\protect\citeauthoryear{Hathaway \etal}{2003}]{hath03}
Hathaway, D. H., Nandy, D., Wilson, R. M., Reichmann, E. J.: 
 \apj\ {\bf 589}, 665 (2003)
\bibitem[\protect\citeauthoryear{Howard}{1996}]{how96}
Howard, R.F.: \araa\ {\bf 34}, 75 (1996)
\bibitem[\protect\citeauthoryear{Hiremath}{2002}]{hi02}
Hiremath, K.M.: \aap\ {\bf 386}, 674 (2002)
\bibitem[\protect\citeauthoryear{{Javaraiah}}{2007}]{jj07}
Javaraiah, J.: \mnras\ {\bf 377}, L34 (2007) 
\bibitem[\protect\citeauthoryear{{Javaraiah}}{2008}]{jj08}
Javaraiah, J.: \solphys\ {\bf 252}, 419 (2008)
\bibitem[\protect\citeauthoryear{{Javaraiah}}{2011}]{jj11}
Javaraiah, J.: \solphys\ {\bf 270}, 463 (2011)
\bibitem[\protect\citeauthoryear{{Javaraiah}}{2012a}]{jj12a}
Javaraiah, J.: \solphys\  {\bf 281}, 827 (2012a)
\bibitem[\protect\citeauthoryear{{Javaraiah}}{2012b}]{jj12b}
Javaraiah, J.: \apss\ {\bf 338}, 217 (2012b)
\bibitem[\protect\citeauthoryear{{Javaraiah}}{2013}]{jj13}
Javaraiah, J.: \solphys\ {\bf 287}, 197 (2013)
\bibitem[\protect\citeauthoryear{{Javaraiah}}{2015}]{jj15}
Javaraiah, J.: \na\ {\bf 34}, 54 (2015)
\bibitem[\protect\citeauthoryear{{Javaraiah and Gokhale}}{1997a}]{jg97a}
Javaraiah, J., Gokhale, M.H.: \solphys\ {\bf 170}, 369 (1997a)
\bibitem[\protect\citeauthoryear{{Javaraiah and Gokhale}}{1997b}]{jg97b}
Javaraiah, J., Gokhale, M.H.: \aap\ {\bf 327}, 795 (1997b)
\bibitem[\protect\citeauthoryear{{Kilcik and Ozguc}}{2014}]{kz14}
Kilcik, A., Ozguc, A.: \solphys\ {\bf 289}, 1379 (2014)
\bibitem[\protect\citeauthoryear{{Kilcik \etal}}{2011}]{klck11}
Kilcik, A., Yurchyshyn, V.B., Abramenko, V., Goode, P. , Ozguc, A., Rozelot,
 J.P., Cao, W.: \apj\ {\bf 731}, 30 (2011)
\bibitem[\protect\citeauthoryear{{Knaack \etal}}{2004}]{knaack04}
Knaack, R., Stenflo, J.O., Berdyugina, S.V.: \aap\ {\bf 418}, L17 (2004)
\bibitem[\protect\citeauthoryear{{Lef\`evre and Clette}}{2011}]{lc11}
Lef\`evre, L.,  Clette, F.A.: 2011, \aap\ {\bf 536}, L11
\bibitem[\protect\citeauthoryear{{Li \etal}}{2002}]{li02}
Li, K.J., Wang, J.X., Xiong, S.Y., Liang, H.F., Yun, H.S.,  Gu, X.M.:  
 \aap\  {\bf 383}, 648 (2002)
\bibitem[\protect\citeauthoryear{{Newton and Milsom}}{1955}]{nm55}
Newton, H.W., Milsom, A. S.:  \mnras\ {\bf 115}, 398 (1955)
\bibitem[\protect\citeauthoryear{{Norton and Gallagher}}{2010}]{ng10}
Norton, A.A.,  Gallagher, J.C.: \solphys\  {\bf 261}, 193 (2010)
\bibitem[\protect\citeauthoryear{{Obridiko and Badalyan}}{2014}]{ob14}
Obridiko, V.N., Badalyan, O.G.: Astron. Rep. {\bf 51}, 936 (2014)
\bibitem[\protect\citeauthoryear{{Ramesh}}{2010}]{ramesh10}
Ramesh, K.B.: \apjl\ {\bf 712}, L77 (2010)
\bibitem[\protect\citeauthoryear{{Ravindra and Javaraiah}}{2015}]{rj15}
Ravindra, B.  Javaraiah, J.: \na\ {\bf 39}, 55 (2015)
\bibitem[\protect\citeauthoryear{{Roy}}{1977}]{roy77}
Roy, J.R.: \solphys\ {\bf 52}, 53 (1977)
\bibitem[\protect\citeauthoryear{{Shapira \etal}}{2011}]{shap11}
Shapiro, A.V., Rozanov, E., Egorova, T., Shapiro, A.I., Peter, Th., 
  Schmutz, W.: J. Atmos. Sol.-Terr. Phys. {\bf 73}, 348 (2011)
\bibitem[\protect\citeauthoryear{Sivaraman \etal}{2003}]{siva03}
Sivaraman, K.R., Sivaraman, H., Gupta, S.S.,  Howard, R.: \solphys\  {\bf 214}, 65 (2003)
\bibitem[\protect\citeauthoryear{{Svalgaard \etal}}{2005}]{sva05}
Svalgaard, L.,  Cliver, E.W.,   Kamide, Y.: \grl\  {\bf 32}, 021664 (2005)
\bibitem[\protect\citeauthoryear{{Swinson \etal}}{1986}]{swin86}
Swinson, D.B.,  Koyama, H.,  Saito, T.: \solphys\ {\bf 106}, 305 (1986)
\bibitem[\protect\citeauthoryear{{Sokoloff and Nesme-Ribes}}{1994}]{snr94}
Sokoloff, D., Nesme-Ribes, E.: \aap\  {\bf 288}, 293 (1994)
\bibitem[\protect\citeauthoryear{{Temmer \etal}}{2006}]{tem06}
Temmer, M., Ryb\'ak, J., Bendik, P., Veronig, A., Vogler, F., Otruba, W.,
 Potzi, W., Hanslmeier, A.: \aap\ {\bf 447}, 735 (2006) 
\bibitem[\protect\citeauthoryear{{Tlatov}}{2015}]{tltv15}
Tlatov, A.G.: Adv. Space Res. {\bf 55}, 851 (2015)
\bibitem[\protect\citeauthoryear{{Verma}}{1993}]{verma93}
Verma, V.K.: \apj\ {\bf 403}, 797 (1993)
\bibitem[\protect\citeauthoryear{Ward}{1965}]{war65}
Ward, F.: \apj\ {\bf 141}, 534 (1965)
\bibitem[\protect\citeauthoryear{Ward}{1966}]{war66}
Ward, F.: \apj\ {\bf 145}, 416 (1966)
\bibitem[\protect\citeauthoryear{{Zolotova and Ponyavin}}{2006}]{zp06}
Zolotova, N.V., Ponyavin, D.I.: \aap\ {\bf 449}, L1 (2006)  
\bibitem[\protect\citeauthoryear{{Zolotova and Ponyavin}}{2014}]{zp14}
Zolotova, N.V., Ponyavin, D.I.: \jgr\ {\bf 119}, 3281 (2014)  
\end{thebibliography}
\end{document}